\documentclass[twocolumn,showpacs,preprintnumbers,amsmath,amssymb,showkeys,nofootinbib]{revtex4}

\usepackage{graphicx}
\usepackage{dcolumn}
\usepackage{bm}
\newcommand{\bqa}{\begin{eqnarray}}
\newcommand{\eqa}{\end{eqnarray}}
\newcommand{\be}{\begin{equation}}
\newcommand{\ee}{\end{equation}}

\begin{document}


\title{Scalar Resonance Contributions to the Dipion Transition Rates of $\Upsilon(4S,5S)$ in the re-scattering model}

\author{Ce Meng$~^{(a)}$ and Kuang-Ta Chao$~^{(a,b)}$}
\affiliation{ {\footnotesize (a)~Department of Physics, Peking
University,
 Beijing 100871, People's Republic of China}\\
{\footnotesize (b)~Center for High Energy Physics, Peking
University, Beijing 100871, People's Republic of China}}


\begin{abstract}

In order to explain the observed unusually large dipion transition
rates of $\Upsilon(10870)$, the scalar resonance contributions in
the re-scattering model to the dipion transitions of $\Upsilon(4S)$
and $\Upsilon(5S)$ are studied. Since the imaginary part of the
re-scattering amplitude is expected to be dominant, the large ratios
of the transition rates of $\Upsilon(10870)$, which is identified
with $\Upsilon(5S)$, to that of $\Upsilon(4S)$ can be understood as
mainly coming from the difference between the $p$-values in their
decays into open bottom channels, and the ratios are estimated
numerically to be about 200-600 with reasonable choices of
parameters. The absolute and relative rates of
$\Upsilon(5S)\to\Upsilon(1S,2S,3S)\pi^+\pi^-$ and
$\Upsilon(5S)\to\Upsilon(1S)K^+K^-$ are roughly consistent with
data. We emphasize that the dipion transitions observed for some of
the newly discovered $Y$ states associated with charmonia may have
similar features to the dipion transitions of $\Upsilon(5S)$.
Measurements on the dipion transitions of $\Upsilon(6S)$ could
provide further test for this mechanism.

\end{abstract}


\maketitle

\section{Introduction}

Hadronic transitions of heavy quarkonia are important for
understanding both the heavy quarkonium dynamics and the formation
of light hadrons. Because heavy quarkonium is expected to be compact
and non-relativistic, at least for the lower-lying states, QCD
multiple expansion (QCDME) approach~\cite{QCDME} can be used in
analysis of these transitions, where the heavy quarkonium system
serves as a compact color source and emits soft gluons which are
hadronized into pions or other mesons.

Applying factorization and using the measurement of $\psi(2S)\to
J/\psi\pi\pi$ as input, the widths of dipion transitions of
$\Upsilon$ system  were successfully predicted~\cite{KuangYan}(see
Ref.~\cite{Kuang06} for an extensive review and the updates; see
also Ref.~\cite{voloshin07} for a comprehensive review on charmonium
hadronic transitions). However, the situation became more
complicated when comparing the predicted $M_{\pi\pi}$ distribution,
which is peaked at the large $M_{\pi\pi}$ region, with the
double-peaked one measured by CLEO~\cite{CLEO82-3S1S,CLEO07-YnSmS}
for $\Upsilon(3S)\to\Upsilon(1S)\pi\pi$. A similar shape was also
found in the $M_{\pi\pi}$ distribution of
$\Upsilon(4S)\to\Upsilon(2S)\pi\pi$ transition~\cite{BaBar06-4S2S}.
Lots of attempts (see~\cite{Kuang06} and references therein) have
been made to improve the QCDME approach. In particular, a study of
the $\Upsilon(4S)$ dipion transitions with $\pi\pi$ interactions was
made in Ref.~\cite{guo}.

More strikingly, the widths of
$\Upsilon(10870)\to\Upsilon(1S,2S)\pi^+\pi^-$ recently measured by
the Belle Collaboration~\cite{Belle07-5SmS} are about 2-3 order in
magnitude larger than those of
$\Upsilon(nS)\to\Upsilon(mS)\pi^+\pi^-$~\cite{BaBar06-4S2S,Belle07_4S1S,PDG06},
where $n=4,3,2$ and $m<n$, with even more complex structures in the
$M_{\pi\pi}$ distributions. If the resonance $\Upsilon(10870)$ is
indeed the $\Upsilon(5S)$ (note that the measured mass and leptonic
width are consistent with this assignment), then its dipion
transitions to lower-lying states can evidently not be described by
the simple multiple expansion approach. The large rates of
$\Upsilon(5S)\to\Upsilon(1S,2S)\pi^+\pi^-$ are puzzling, and new
mechanisms seem to be needed to explain them. (In this paper we will
focus on the possibility that $\Upsilon(10870)$ is the
$\Upsilon(5S)$, and leave discussions on other possible assignments
e.g. $b\bar bg$ hybrids or $b\bar bq\bar q$ tetraquarks for
$\Upsilon(10870)$ (see, e.g.\cite{hou}) elsewhere.)

In general, for higher excited states of charmonium and bottomonium,
the radius becomes larger, and can even be larger than the range of
the soft gluon field if they are high enough. Then, for these exited
heavy quarkonia, justification of QCDME scenario becomes
problematic.  Particularly, when the excited state lies above the
open flavor thresholds, the coupled-channel effects will change the
QCDME scenario markedly and add new mechanisms to the analysis of
its dipion transitions. Some of these effects were studied in
Ref.~\cite{Zhou91}, but the effects were found to be tiny for
$\Upsilon(3S,2S)\to\Upsilon(2S,1S)\pi\pi$. This is probably due to
the fact that $\Upsilon(3S,2S)$ are too far below the open flavor
threshold of $B\bar{B}$. However, the case should be changed for
$\Upsilon(5S)$ and $\Upsilon(4S)$, since open bottom channels, such
as $B^{(*)}\bar{B}^{(*)}$ and $B_s^{(*)}\bar{B}_s^{(*)}$, can be
open and contribute to their transition rates significantly.

One important feature of coupled decay channels for $\Upsilon(5S)$
and $\Upsilon(4S)$ is the final state interaction, i.e., in the
decay the $B^{(*)}$ and $\bar{B}^{(*)}$ can interact with each other
at long distances and then convert into a lower $\Upsilon$ plus
light mesons. For simplification, in this paper, we will use the
re-scattering
model~\cite{Cheng05_ReSC,Liu07_X3872_ReSC,Meng07_X3872_ReSC,Meng07_Z4430_ReSC}
to study the scalar resonance contributions to the dipion transition
rates of $\Upsilon(5S)$ as well as $\Upsilon(4S)$. In this picture,
the higher $\Upsilon$ decays into $B^{(*)}\bar{B}^{(*)}$ first, and
then through one $B^{(*)}$ meson exchange turns into another lower
$\Upsilon$ and a scalar resonance, such as $\sigma$ or $f_0(980)$
(perhaps also $f_0(1370)$), which couples to the dipion.
Experimentally, for both charmonium dipion transitions (see, e.g.,
\cite{bes07}) and bottomonium dipion transitions (see,
\cite{CLEO07-YnSmS}), the dipion systems are found to be dominated
by the S-wave, therefore the scalar resonances could play an
essential role in these transitions. In fact, the scalar resonance
i.e. the $\sigma$ dominance approach has been used to fit the
$\psi(2S)$ data\cite{bes07}. In addition, including the
contributions from scalar resonances could be helpful to explain the
$M_{\pi\pi}$ distributions in
$\Upsilon(3S,2S)\to\Upsilon(2S,1S)\pi\pi$
~\cite{Komada01-scalar-contributions,Uehara03-scalar-contributions},
especially the double-peaked structure of mass distribution in
$\Upsilon(3S)\to\Upsilon(1S)\pi\pi$. It would also be interesting to
further examine the complex $M_{\pi\pi}$ distributions in the dipion
transitions of $\Upsilon(4S)$~\cite{BaBar06-4S2S} and
$\Upsilon(5S)$~\cite{Belle07-5SmS}.

In this paper we will assume that in the $\Upsilon(4S,5S)$ dipion
transitions the two pions are produced mainly via scalar resonances
coupled  to intermediate $B^{(*)}$ mesons due to the long-distance
final state interactions. We will discuss the model and calculate
the transition rates of $\Upsilon(4S,5S)$. A summary will be given
in the last section.

\section{The model}

\begin{figure}[t]
\begin{center}
\vspace{0cm}
 \hspace*{0cm}
\scalebox{0.5}{\includegraphics[width=16cm,height=13cm]{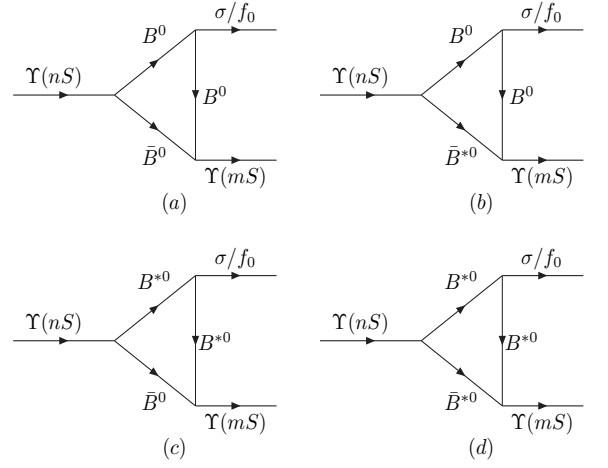}}
\end{center}
\vspace{0cm}\caption{The diagrams for $\Upsilon(nS)\to
B^{(*)0}\bar{B}^{(*)0}\to\Upsilon(mS)\mathcal{S}$.}\label{Fig:Y-YS}
\end{figure}

In the re-scattering model, the transitions
$\Upsilon(4S,5S)\to\Upsilon(1S,2S)\!\!\!~\mathcal{S}$ can arise from
scattering of intermediate state $B^{(*)}\bar{B}^{(*)}$ by exchange
of another $B$ meson. Here, $\mathcal{S}$ denotes scalar resonance
$\sigma$ or $f_0(980)$ (perhaps also $f_0(1370)$, which will decay
to $\pi\pi(K\bar{K})$ eventually. The typical diagrams are shown in
Fig.~\ref{Fig:Y-YS}, and the other ones can be related to those in
Fig.~\ref{Fig:Y-YS} by charge conjugation transformation
$B\leftrightarrow\bar{B}$ and isospin transformation
$B^0\leftrightarrow B^+$ and $\bar{B}^0\leftrightarrow B^-$.
Therefore, the amplitudes of Fig.~\ref{Fig:Y-YS}(a,b,c,d) should be
multiplied by a factor of 4, respectively.

To evaluate the amplitudes, we need the following effective
Lagrangians:
\begin{subequations} \label{effective-Lagrangians}
\begin{eqnarray}
\mathcal{L}_{\Upsilon BB}&=& g_{\Upsilon
BB}\Upsilon_\mu(\partial^\mu
B{B}^{\dagger}-B\partial^\mu {B}^{\dagger}),\label{L-YBB}\\
\mathcal{L}_{\Upsilon B^*B}&=& \!\frac{g_{\Upsilon\!
B^*\!B}}{m_{\Upsilon}}\varepsilon^{\mu\nu\alpha\beta}\partial_\mu
\!\Upsilon_\nu
\!\nonumber\\
&& \times(B^*_\alpha\overleftrightarrow{\partial}_\beta
{B}^{\dagger}\!\! - \!\!
B\overleftrightarrow{\partial}_\beta{B}^{*\dagger}_\alpha\!),\label{L-YB*B}\\
\mathcal{L}_{\Upsilon B^*B^*}&=& g_{\Upsilon B^* B^*} (
-\Upsilon^\mu
B^{*\nu}\overleftrightarrow{\partial}_\mu {B}_\nu^{*\dagger} \nonumber\\
&&+ \Upsilon^\mu B^{*\nu}\partial_\nu{B}^{*\dagger}_{\mu} -
\Upsilon_\mu\partial_\nu B^{*\mu}
{B}^{*\nu\dagger}),\label{L-YB*B*}\\
\mathcal{L}_{\mathcal{S} BB}&=& g_{\mathcal{S} BB}\mathcal{S}
B{B}^{\dagger},\label{L-SBB}\\
\mathcal{L}_{\mathcal{S} B^*B^*}&=& -g_{\mathcal{S}
B^*B^*}\mathcal{S} B^*\cdot{B^*}^{\dagger},\label{L-SBB}
\end{eqnarray}
\end{subequations}
where
$\overleftrightarrow{\partial}=\overrightarrow{\partial}-\overleftarrow{\partial}$.
In the heavy quark limit, the coupling constants in
(\ref{effective-Lagrangians}) can be related to each other by heavy
quark symmetry as: \bqa g_{\Upsilon BB}&=&g_{\Upsilon
B^*B}=g_{\Upsilon B^*B^*}\,\label{HQS:g-YBB}\\
g_{\mathcal{S} BB}&=&g_{\mathcal{S} B^*B^*}.\label{HQS:g-SBB} \eqa
Particularly, the coupling constants for $\Upsilon(4S)$ and
$\Upsilon(5S)$ can be determined by the observed values of their
partial decay widths.

All the coupling constants will be determined in the next section.
However, it is necessary to emphasize here that the determinations
will not account for the off-shell effect of the exchanged $B^{(*)}$
meson, of which the virtuality can not be ignored. Such effects can
be compensated by introducing, e.g., the
monopole~\cite{Cheng05_ReSC} form factors for off-shell vertexes.
Let $q$ denote the momentum transferred and $m_i$ the mass of
exchanged meson, the form factor can be written as
\begin{eqnarray}\label{formfactor1}
\mathcal{F}(m_{i},q^2)=\frac{(\Lambda+m_i)^{2}-m_{i}^2
}{(\Lambda+m_i)^{2}-q^{2}}.
\end{eqnarray}
We will fix the cutoff $\Lambda=660$ MeV~\cite{Meng07_X3872_ReSC} in
our numerical analysis in the next section.

As emphasized in Ref.~\cite{Meng07_Z4430_ReSC}, the form factor
suppression favors the production of higher-excited heavy quarkonium
state over that of the lower one because the mass of the former is
closer to the open flavor threshold than the later. This effect will
largely balance the final-state phase space factor, which favors the
production of lower state, and will give a reasonable relative rate
between $\Upsilon(5S)\to\Upsilon(1S)\pi\pi$ and
$\Upsilon(5S)\to\Upsilon(2S)\pi\pi$ transitions, as one can see in
the next section.

We are now in a position to compute the contributions of diagrams in
Fig.~\ref{Fig:Y-YS}.  If the $\Upsilon(nS)$ state lies above the
$B^{(*)}\bar{B}^{(*)}$ threshold, the absorptive part (imaginary
part) of the amplitude arising from Fig.~\ref{Fig:Y-YS} can be
evaluated by the Cutkosky rule. For the process $\Upsilon(nS)\to
B^{(*)}(p_1)+\bar{B}^{(*)}\to \Upsilon(mS)+\mathcal{S}$, the
absorptive part of the amplitude reads
\bqa\label{Abs:CutRule}
\textbf{Abs}_i&=&\frac{|\vec{p}_1|}{32\pi^2m_{\Upsilon(nS)}}\int
d\Omega
\mathcal{A}_i(\Upsilon(nS)\to B^{(*)}\bar{B}^{(*)})\nonumber\\
&&\times \mathcal{C}_i(B^{(*)}\bar{B}^{(*)}\to
\Upsilon(mS)\mathcal{S}), \eqa
where $i=(a,b,c,d)$, and $d\Omega$ and $\vec{p}_1$ denote the solid
angle of the on-shell $B^{(*)}\bar{B}^{(*)}$ system and the
3-momentum of the on-shell $B^{(*)}$ meson in the rest frame of
$\Upsilon(nS)$, respectively. Constrained by the heavy quark
symmetry relation (\ref{HQS:g-YBB}) and (\ref{HQS:g-SBB}), the
amplitude of Fig~\ref{Fig:Y-YS}(b) is equal to that of
Fig~\ref{Fig:Y-YS}(c) in the heavy quark limit. Therefore, the total
absorptive part of the re-scattering amplitude can be given by
\be\label{Abs:total}
\textbf{Abs}\thickapprox4\textbf{Abs}_a+8\textbf{Abs}_b+4\textbf{Abs}_d.
\ee

The evaluation of the real part of the amplitude is difficult to be
achieved, and will bring large uncertainties inevitably.
Fortunately, for the transitions
$\Upsilon(4S,5S)\to\Upsilon(1S,2S)\mathcal{S}$, the contributions
from the real part are expected to be small, because the masses of
$\Upsilon(4S,5S)$ are not very close to the open flavor thresholds
as those of $X(3872)$~\cite{Meng07_X3872_ReSC} and
$Z(4430)$~\cite{Meng07_Z4430_ReSC}. Thus we assume that the
contributions from the real part can be neglected, and use
(\ref{Abs:CutRule}) to determine the full amplitude in the
calculations.

In the absorptive part, the intermediate states
$B^{(*)}\bar{B}^{(*)}$ are on-shell, and the amplitude in
(\ref{Abs:CutRule}) is proportional to the phase space factor of
decay $\Upsilon(nS)\to B^{(*)}\bar{B}^{(*)}$:
\be
\frac{|\vec{p}_1|}{32\pi^2m_{\Upsilon(nS)}}\label{Phase-space}.\ee
The amplitude $\mathcal{A}_i$ in (\ref{Abs:CutRule}) is also
proportional to $|\vec{p}_1|$ since it involves an on-shell P-wave
vertex $\Upsilon(nS) B^{(*)}B^{(*)}$. Furthermore, a hidden factor
$|\vec{p}_1|$ will emerge after performing the integral in
(\ref{Abs:CutRule}) explicitly. As a result, the amplitude
$\textbf{Abs}_i$ is proportional to $|\vec{p}_1|^3$. This fact is
important in understanding the huge difference between the decay
rates of $\Upsilon(5S)\to\Upsilon(1S,2S)\pi\pi$ and
$\Upsilon(4S)\to\Upsilon(1S,2S)\pi\pi$, since for $\Upsilon(4S)$
only one decay channel $B\bar B$ is really open, and the
corresponding $p$-value $|\vec{p_1}|$ is small, whereas for
$\Upsilon(5S)\to\Upsilon(1S,2S)\pi\pi$ three decay channels $B\bar
B, B^*\bar B+c.c., B^*\bar B^*$ are all open with rather large
$p$-values. Similar to this, the essential role played by the phase
space factor in determining the absorptive part of the re-scattering
amplitude has been emphasized in the case of X(3872)in
Refs.\cite{Liu07_X3872_ReSC,Meng07_X3872_ReSC}, where the tiny phase
space greatly suppresses the absorptive part of the re-scattering
amplitude.

Generally, to compare the result with the experimental measurement,
one needs to describe the transition amplitude to $\pi\pi(K\bar{K})$
for a virtual scalar resonance $\mathcal{S}$ explicitly. However,
since we will focus on the total rate,  we treat $\mathcal{S}$ as
narrow resonance and use Breit-Wigner distribution
\be \mathcal{F}_{\mathcal{S}}(t)=\frac{1}{\pi}\frac{\sqrt
t\Gamma_{\mathcal{S}}(t)}{(t-m_{\mathcal{S}}^2)^2+m_{\mathcal{S}}\Gamma_{\mathcal{S}}(t)^2}\label{BW-distribution}\ee
to describe the resonance in the calculation of cross sections, as
the treatment of $\rho$ resonance in Ref.~\cite{Meng07_X3872_ReSC}.
In (\ref{BW-distribution}), the variable $t$ denotes the momentum
squared of $\mathcal{S}$, and the function $\Gamma_{\mathcal{S}}(t)$
is given by
\bqa
&&\Gamma_{\mathcal{S}}(t)=\frac{p_{\pi}g_{\mathcal{S}\pi\pi}}{8\pi
t}+\frac{p_{K}g_{\mathcal{S}KK}}{8\pi t},\label{Total-width:scalar}\\
&&p_{\pi}=\sqrt{\frac{t}{4}-m_{\pi}^2},~~~p_{K}=\sqrt{\frac{t}{4}-m_{K}^2}\nonumber\eqa
The resonance parameters in (\ref{BW-distribution}) and the coupling
constants in (\ref{Total-width:scalar}) will be evaluated in the
next section following Ref.~\cite{Komada01-scalar-contributions}.

\section{Numerical Results and Discussions}
Since the contribution from the absorptive part of the re-scattering
amplitude corresponds to the real decay process $\Upsilon(nS)\to
B^{(*)}\bar{B}^{(*)}$, the coupling constants
$g_{\Upsilon(nS)B^{(*)}B^{(*)}}$ should be determined by the
measured values of the decay widths of $\Upsilon(4S,5S)\to
B^{(*)}\bar{B}^{(*)}$~\cite{PDG06}, and the results are given by
\bqa g_{\Upsilon(4S)BB}&=& 24,\label{g:Upsilon(4S)BB}\\
 g_{\Upsilon(5S)BB}&<& 2.9,\label{g:Upsilon(5S)BB}\\
 g_{\Upsilon(5S)B^*B}&=& 1.4\pm0.3,\label{g:Upsilon(5S)B*B}\\
 g_{\Upsilon(5S)B^*B^*}&=& 2.5\pm0.4.\label{g:Upsilon(5S)B*B*}\eqa
The value of $g_{\Upsilon(4S)BB}$ in (\ref{g:Upsilon(4S)BB}) is
typical, and is comparable to the estimation using the vector meson
dominance model~\cite{Meng07_X3872_ReSC} for $g_{\Upsilon(1S)BB}$:
\be g_{\Upsilon BB}\thickapprox
\frac{m_{\Upsilon(1S)}}{f_{\Upsilon(1S)}}\sim
15,\label{g:UpsilonBB:VMD}\ee
where the decay constant $f_{\Upsilon(1S)}$ can be determined by the
leptonic width of $\Upsilon(1S)$. However, the values determined
from the $\Upsilon(5S)$ data in
(\ref{g:Upsilon(5S)BB})-(\ref{g:Upsilon(5S)B*B*}) are small. This
may be partly due to the fact that as a high-excited $b\bar{b}$
state, the wave function of $\Upsilon(5S)$ has a complicated node
structure, and the coupling constants will be small if the
$p$-values of $B^{(*)}\bar{B}^{(*)}$ channels (1060-1270 MeV) are
close to those corresponding to the zeros in the amplitude. The
symmetry relation in (\ref{HQS:g-YBB}) can also be violated by the
same reason.

As for the coupling constants $g_{\Upsilon(mS)B^{(*)}B^{(*)}}$
($m<5$), we assume that the symmetry relations in (\ref{HQS:g-YBB})
hold, and they are equal to each other, which is implied by
comparison between (\ref{g:Upsilon(4S)BB}) and
(\ref{g:UpsilonBB:VMD}).

Numerically, we find that the amplitude $\textbf{Abs}_a$ is
relatively small, and the the amplitude $\textbf{Abs}_b$ partly
cancels $\textbf{Abs}_d$ in (\ref{Abs:total}). So we choose
$g_{\Upsilon(5S)BB}= 2.5$, and focus on the sensitivities of the
decay rates to the coupling constants $g_{\Upsilon(5S)B^*B}$ and
$g_{\Upsilon(5S)B^*B^*}$ in (\ref{g:Upsilon(5S)B*B}) and
(\ref{g:Upsilon(5S)B*B*}).

The phenomenological coupling constants
$g_{\mathcal{S}B^{(*)}B^{(*)}}$ are difficult to be determined.
However, in the linear realization of chiral symmetry, one can
relate them to the coupling constant~\cite{HMchiEL}
\be g_{B^*B\pi}= \frac{2g m_{B}}{f_\pi},\ee
where $g\thickapprox0.6$~\cite{Meng07_X3872_ReSC}, and $f_\pi$ is
the decay constant of $\pi$. In general, the coupling constant
$g_{\mathcal{S}B^{(*)}B^{(*)}}$ could be obtained through scaling
$g_{B^*B\pi}$ by a typical chiral scale like $f_\pi$. Thus, they are
of order $\mathcal{O}(m_B)$, and we choose
\bqa g_{\sigma BB}&=& g_{\sigma B^*B^*}= 10~\mbox{GeV},\label{g:sigmaBB}\\
 g_{f0 BB}&=& g_{f0 B^*B^*}= 10\sqrt{2}~\mbox{GeV}.\label{g:f0BB}\eqa
Here, in (\ref{g:f0BB}) we introduce a numerical factor of
$\sqrt{2}$, which is somewhat arbitrary,  to roughly account for the
contributions from other higher scalar resonances, such as
$f_0(1370)$.

In the linear realization of chiral symmetry, there should exist
coupling of $BB\pi\pi$, which will cancel the one of $BB\sigma$ in
the low-energy limit. However, the cancelation is no longer
effective in processes with large energy
release~\cite{Komada01-scalar-cancelation}.   We will not take into
account this cancelation in the present paper and leave it to be
studied in the future.

The scalar resonance parameters, which are listed in
Tab.~\ref{Tab:Scalar-resonance}, are chosen following
Ref.~\cite{Komada01-scalar-contributions} (while $m_{f_0(980)}$
following Ref.~\cite{PDG06}), where they are determined by fitting
the $M_{\pi\pi}$ distributions in
$\Upsilon(2S,3S)\to\Upsilon(1S,2S)\pi\pi$, $\psi(2S)\to
J/\psi\pi\pi$ and $J/\psi\to\phi\pi\pi(KK)$.

\begin{widetext}
\begin{center}\begin{table}
\caption{Resonance parameters of $\sigma$ and
$f_0(980)$~\cite{Komada01-scalar-contributions,PDG06}.}

\begin{tabular}{cccccc}
\hline
   & $m_\mathcal{S}$(MeV) & $g_{\mathcal{S}\pi\pi}$(GeV) &
   $\Gamma_{\mathcal{S}\pi\pi}$(MeV)
                 & $g_{\mathcal{S} KK}$(GeV) & $\Gamma_{\mathcal{S} KK}$/MeV  \\
\hline
$\sigma$   & $526\pm30$ & 3.06 & $302\pm10$ &       &          \\
$f_0(980)$ & $980\pm10$ & 1.77 & $61\pm1$   & 2.70  & $12\pm1$ \\
\hline \label{Tab:Scalar-resonance}\end{tabular}

\end{table}\end{center}

\begin{center}\begin{table}
\caption{Transition widths of
$\Upsilon(nS)\to\Upsilon(mS)\pi^+\pi^-/K^+K^-$ in units of KeV.
Respectively, the contributions from $\sigma$ and $f_0(980)$ are
listed in the second and the third columns, and the error bars come
from those of $m_\sigma$($m_{f_0(980)}$), $g_{\Upsilon(5S)B^*B^*}$
and $g_{\Upsilon(5S)B^*B}$ in turn. Experimental data of
$\Upsilon(4S)\to\Upsilon(1S,2S)\pi^+\pi^-$ are taken from
Ref.~\cite{BaBar06-4S2S}, and the others from
Ref.~\cite{Belle07-5SmS}.}
\begin{tabular}{ccccc}
\hline
   & from $\sigma$ & from $f_0(980)$ &
   total           & Experimental data  \\
\hline
$\Upsilon(4S)\to\Upsilon(1S)\pi^+\pi^-$   & $0.54^{+0.00}_{-0.00}$ & $0.93^{+0.03}_{-0.03}$ &  $1.47\pm0.03$  & $1.8\pm0.4$         \\
$\Upsilon(4S)\to\Upsilon(2S)\pi^+\pi^-$   & $1.09^{+0.23}_{-0.21}$ & $0.05^{+0.00}_{-0.01}$ &  $1.14^{+0.23}_{-0.21}$  & $2.7\pm0.8$  \\
\hline\hline $\Upsilon(5S)\to\Upsilon(1S)\pi^+\pi^-$  &
$102^{+1+42+21}_{-0-35-9}$ & $225^{+1+93+47}_{-1-77-43}$ &
$327^{+114}_{-97}$ & $590\pm 40 \pm 90$\\
$\Upsilon(5S)\to\Upsilon(2S)\pi^+\pi^-$  &
$385^{+10+164+87}_{-11-135-78}$ & $37^{+4+16+9}_{-3-13-7}$ &
$422^{+187}_{-157}$ & $850\pm 70 \pm 160$\\
$\Upsilon(5S)\to\Upsilon(3S)\pi^+\pi^-$  &
$306^{+78+133+73}_{-64-108-64}$ & $13^{+1+6+4}_{-1-4-2}$ &
$319^{+171}_{-141}$ & $520^{+200}_{-170}\pm 100$\\
\hline\hline $\Upsilon(5S)\to\Upsilon(1S)K^+K^-$  &
 & $32^{+5+13+5}_{-5-11-6}$ &
$32^{+15}_{-13}$ & $67^{+17}_{-13}\pm 13$\\
\hline \label{Tab:Predictions-Width}\end{tabular}
\end{table}\end{center}
\end{widetext}

Neglecting the interference between contributions from $\sigma$ and
that from $f_0(980)$, we can now evaluate the transition widths
$\Upsilon(4S,5S)\to\Upsilon(1S,2S)\pi^+\pi^-$ using the parameters
and the coupling constants determined above, and the results are
listed in Tab.~\ref{Tab:Predictions-Width}. Since the transitions
$\Upsilon(5S)\to\Upsilon(3S)\pi^+\pi^-$ and
$\Upsilon(5S)\to\Upsilon(1S)K^+K^-$ are also
observed~\cite{Belle07-5SmS} with large rates and quite high
statistic significance ($3.2\sigma$ and $4.9\sigma$, respectively),
we also evaluate the corresponding rates to be compared with the
experimental data. Contributions from $\sigma$ and $f_0(980)$ are
listed in the second and third columns in
Tab.~\ref{Tab:Predictions-Width}, respectively. The error bars in
these two columns come from those of $m_\sigma$($m_{f_0(980)}$),
$g_{\Upsilon(5S)B^*B^*}$ and $g_{\Upsilon(5S)B^*B}$ in turn, and the
signs "$+$" correspond the smaller $m_\sigma$($m_{f_0(980)}$), the
larger $g_{\Upsilon(5S)B^*B^*}$ and the smaller
$g_{\Upsilon(5S)B^*B}$, respectively. The only exception is that the
width of $\Upsilon(5S)\to\Upsilon(1S)K^+K^-$ increases with
$m_{f_0(980)}$. The sensitivity of the width of
$\Upsilon(5S)\to\Upsilon(3S)\pi^+\pi^-$ to the parameter $m_\sigma$
can be easily understood since the phase space is small and can only
cover part of the distribution of $\sigma$ resonance. On the other
hand, the sensitivities of the widths of
$\Upsilon(5S)\to\Upsilon(1S,2S,3S)\pi^+\pi^-$ and
$\Upsilon(5S)\to\Upsilon(1S)K^+K^-$ to the coupling constant
$g_{\Upsilon(5S)B^*B}$ are mainly due to the partial cancelation
between amplitudes $\textbf{Abs}_b$ and $\textbf{Abs}_d$.

The dependence on the cutoff $\Lambda$, which is defined in
(\ref{formfactor1}), is not shown in
Tab.~\ref{Tab:Predictions-Width}. In our evaluations, we have chosen
$\Lambda=660$ MeV following Ref.~\cite{Meng07_X3872_ReSC}. If the
cutoff increases(decreases) by, say, 220 MeV, the rates listed in
Tab.~\ref{Tab:Predictions-Width} will increase(decrease) by 2-3
times in magnitude correspondingly.

As mentioned in last section, the main difference between
$\Upsilon(5S)\to\Upsilon(1S,2S,3S)\pi\pi$ and
$\Upsilon(4S)\to\Upsilon(1S,2S)\pi\pi$ in this re-scattering model
is the number of open flavor channels involved and, essentially, the
corresponding $p$-values $|\vec{p_1}|$. Especially, the contribution
to the rate from a given channel is proportional to $|\vec{p_1}|^6$,
which can cause a big difference between the partial widths of
$\Upsilon(5S)$ and $\Upsilon(4S)$ of order
$\mathcal{O}(10^3\mbox{-}10^4)$ in magnitude. After other
ingredients, such as the difference between coupling constants in
(\ref{g:Upsilon(4S)BB}-\ref{g:Upsilon(5S)B*B*}), are taken into
account, the difference becomes about 200-600 in magnitude and,
although with large error bars, is in rough agreement with
experimental data.

The above results are obtained with the assumption of the absorptive
part dominance. If the real part of the re-scattering amplitude can
not be neglected, the difference between the transition decay widths
of $\Upsilon(5S)$ and $\Upsilon(4S)$ will decrease, since the
contributions from the real part do not obey the $|\vec{p_1}|^6$
rule. To clarify to what extent the absorptive part dominance
assumption is sensible, we evaluated the real part by using the
dispersion relation~\cite{Meng07_X3872_ReSC}. To evaluate the
dispersive integral e.g. for the $B\bar B$ decay channel, we take
the upper limit of the integral to be $s_{max}=(m_B+m_B+\Delta)^2$
and choose the cutoff $\Delta$ to be equal to the splitting
$m_{B^*}-m_{B}$ following Ref.~\cite{Meng07_X3872_ReSC}. This choice
of the cutoff will
lead to a contribution to the rates of less than 1 KeV  for all
modes that we are interested in.
However, if one chooses the cutoff $\Delta=100$ MeV, the
contributions of the real part will increase and result in rates of
about 10 KeV for $\Upsilon(4S)\to\Upsilon(1S,2S)\pi\pi$ decays. So,
the $\Upsilon(4S)$ decays are sensitive to the cutoff in the real
part. Nevertheless, this sensitivity does not affect the calculated
large difference between the transition widths of $\Upsilon(5S)$ and
$\Upsilon(4S)$, which is the main point addressed in this paper. In
addition, the above results are obtained by using the couplings
shown in (10)-(13), and if we choose a smaller coupling of
$g_{\Upsilon(4S)BB}$ (say, to be equal to $g_{\Upsilon(5S)BB}$),
with other parameters readjusted, then the real part contribution to
$\Upsilon(4S)$ transitions will be much less than 1 KeV even with
the cutoff $\Delta=100$ MeV. So, despite of large uncertainties with
the model and chosen parameters, the absorptive part dominance
should be a reasonable assumption unless the resonance is very close
to the open channel threshold (as in the case of
X(3872)~\cite{Meng07_X3872_ReSC}, whose mass departs from the $DD^*$
threshold by less than a few MeV). In any case, however, a more
reliable approach for estimating the real part contribution in the
re-scattering model is needed and deserves further study.


As we have mentioned in last section, in the $\Upsilon(5S)$ decays
the $\Upsilon$ form factor defined in (\ref{formfactor1}) favors the
production of higher $\Upsilon$ state over the lower one, while the
final-state phase space plays an opposite role. Thus, as one can see
in Tab.~\ref{Tab:Predictions-Width}, the contributions from $\sigma$
favor the production of $\Upsilon(2S,3S)$ since the phase space
differences are relatively small, while those from $f_0(980)$
(perhaps also $f_0(1370)$) favor the production of $\Upsilon(1S)$.
This fact is important to obtaining a reasonable ratio between the
widths of $\Upsilon(5S)\to\Upsilon(1S)\pi^+\pi^-$ and
$\Upsilon(5S)\to\Upsilon(2S)\pi^+\pi^-$, and may also imply that the
$M_{\pi\pi}$ spectrum of $\Upsilon(5S)\to\Upsilon(1S)\pi^+\pi^-$
should be concentrated in higher mass regions, which is in agreement
with the experimental measurement~\cite{Belle07-5SmS}.

The $M_{\pi\pi}$ spectrum of $\Upsilon(5S)\to\Upsilon(1S)\pi^+\pi^-$
in our model is also dependent on the ratio of $g_{f_0 BB}$ to
$g_{\sigma BB}$, which is assumed to be $\sqrt{2}$ from
(\ref{g:sigmaBB}) and (\ref{g:f0BB}). If we  choose a relatively
larger value for this ratio, the spectrum will be even more
concentrated to higher masses. However, because the ratio is
introduced to account for the contributions from higher scalar
resonances in a rather arbitrary way, and the non-resonance
contributions are neglected in our model, we are unable to fit the
$M_{\pi\pi}$ spectrum at the quantitative level.

We can make a rough prediction for the dipion transitions of
$\Upsilon(11020)$, if it is identified with the $\Upsilon(6S)$.
Constraining the coupling constant $g_{\Upsilon B^{(*)}B^{(*)}}$ by
the total width of $\Upsilon(11020)$, we  find that the dominant
transition mode of $\Upsilon(11020)$ could be $\Upsilon(3S)\pi\pi$
with a large partial width of about 1-2 MeV, while that for
$\Upsilon(1S)\pi\pi$ is about 300 KeV. Of course, due to the large
uncertainties from model parameters, this estimate only serves as a
possible tendency that $\Upsilon(3S)\pi\pi$ and $\Upsilon(2S)\pi\pi$
will be favored over $\Upsilon(1S)\pi\pi$ in the $\Upsilon(6S)$
decays.

The situation in the $c\bar{c}$ system can be similar to but more
complicated than the $b\bar{b}$ system. In fact in the ISR (Initial
State Radiation) process a number of $Y$-states have been
found~\cite{Babar-Ystate,Belle-Ystate} to decay to $J/\psi$ or, but
not "and", $\psi(2S)$ through dipion transitions with large decay
rates, while there seem no "adequate" assignments in the
conventional charmonium family for these states. However, the
abnormal large rates of dipion transitions of these $Y$-states might
indicate that they, or at least some of them, are indeed the
conventional charmonium states that are coupled to the open charm
meson channels, just as the $\Upsilon(10870)$ in the $b\bar b$
system discussed above.  The coupled channel effects are expected to
be more complicated for higher $c\bar c$ states than for $b\bar b$,
since more open charm channels, e.g. $D_1\bar{D}$, aside from
$D^{(*)}\bar D^{(*)}$, are involved for higher charmonia. The large
rates of dipion transitions of these charmonium states might be
accounted for in the re-scattering picture. The
fact~\cite{Belle-Ystate} that the lower $Y$-states are only found in
$J/\psi\pi\pi$ mode while the higher ones in $\psi(2S)\pi\pi$ might
be understood as signals of the competition between the form factor,
which favors $\psi(2S)$, and the final-state phase space, which
favors $J/\psi$, in the re-scattering model, especially when the
contributions from resonances, such as $\sigma$ and $f_0(980)$
(perhaps also $f_0(1370)$), are dominant.

\section{Summary}

In summary, we study the long-distance final state interactions in
$\Upsilon(4S,5S)$ dipion transitions. We calculate the scalar
resonance contributions in the re-scattering model to the dipion
transition rates of $\Upsilon(4S)$ and $\Upsilon(10870)$, which is
identified with $\Upsilon(5S)$, in order to explain the observed
unusually large dipion transition rates of $\Upsilon(10870)$. We
assume that the $\Upsilon(4S,5S)$ decay into $B^{(*)}\bar{B}^{(*)}$
first, and then through one $B^{(*)}$ meson exchange turn into
another lower $\Upsilon$ and a scalar resonance, such as $\sigma$ or
$f_0(980)$ (perhaps also $f_0(1370)$), which couples to the dipion.
Assuming the imaginary part of the re-scattering amplitude
dominates, the large ratios of the transition rates of
$\Upsilon(5S)$ to that of $\Upsilon(4S)$ can be understood as mainly
coming from the difference between the $p$-values of their decays
into open bottom channels, and are estimated to be about 200-600
with reasonable choices for the parameters. Besides, the absolute
and relative rates of $\Upsilon(5S)\to\Upsilon(1S,2S,3S)\pi^+\pi^-$
and $\Upsilon(5S)\to\Upsilon(1S)K^+K^-$ are also roughly compatible
with experimental data. We find that the competition between the
form factor and the final-state phase space plays an important role
in the determination of these relative rates and the $M_{\pi\pi}$
distribution in $\Upsilon(5S)\to\Upsilon(1S)\pi^+\pi^-$. We
emphasize that the dipion transitions observed for some of the newly
discovered $Y$ states associated with charmonia may have similar
features to the dipion transitions of $\Upsilon(5S)$ discussed here.
Measurements on the dipion transitions of $\Upsilon(6S)$ could
provide further test for this mechanism.

Recently, in Ref.\cite{Simonov07-nSmS} the author suggested an
approach  to explain the dipion transitions of $\Upsilon(4S,5S)$,
which is similar, in some sense, to ours but without introducing
scalar resonances. It will be interesting to compare our result with
theirs when their result for $\Upsilon(4S,5S)$ comes out.

\begin{acknowledgments}
We wish to thank H.Q. Zheng for helpful discussions. We also thank
H.Y. Cheng and W.S. Hou for useful comments at the 5th Workshop on
Heavy Flavor Physics and CP Violation, Sanya, China, Dec. 15-19,
2007, where this result was reported. This work was supported in
part by the National Natural Science Foundation of China (No
10421503, No 10675003, No 10721063), and the Research Found for
Doctorial Program of Higher Education of China.
\end{acknowledgments}

\bibliography{apssamp}

\end{document}